\documentclass[reprint,
 amsmath,amssymb,
 aps,prl,
]{revtex4-1}
\usepackage{cancel}
\usepackage{graphicx}
\usepackage{dcolumn}
\usepackage{bm}
\usepackage{physics}
\usepackage{xcolor}
\usepackage{bbm}
\usepackage{textcomp}
\usepackage{natbib}

\DeclareMathOperator\arctanh{arctanh}

\renewcommand{\figurename}{\textbf{Fig.}}
\begin{document}

\title{Anomalous response in the orbital magnetic susceptibility of 2D topological systems}

\author{Daniel Fa\'ilde }

\affiliation{Departamento de Física Aplicada, Instituto de Investigacións Tecnoloxicas,Universidade de Santiago de Compostela,E-15782 Campus Vida s/n, Santiago de Compostela, Spain}

\author{Daniel Baldomir}

\affiliation{Departamento de Física Aplicada, Instituto de Investigacións Tecnoloxicas,Universidade de Santiago de Compostela,E-15782 Campus Vida s/n, Santiago de Compostela, Spain}


\begin{abstract}
Two-dimensional compounds with non-zero Berry curvature are ideal systems to study exotic and technologically favourable thermoelectric and magnetoelectric properties. Within this class of materials, the topological trivial and non-trivial regimes had to present very different behaviours which are encoded for the orbital susceptibility and magnetization.  In order to try to reveal them, we have found that it was necessary to introduce a k-dependent mass term in the relativistic formalism of these materials. Thus, while a topologically trivial insulator is predicted to have a very limited response, in the non-trivial regime we unveil a singular contribution to the orbital magnetic susceptibility which is inversely proportional to the square of the quantum magnetic flux. In this emergent scenario, besides determining the measurement conditions we also find a new route for enhancing the intrinsic orbital magnetism of topological materials widening the range of temperatures and magnetic fields without involving tiny band gaps.
\end{abstract}

\maketitle


The study of the thermoelectric and magnetoelectric properties in non-zero Berry curvature compounds have gathered great interest during the last years \cite{PhysRevB.78.195424,Baldomir2019,RevModPhys.82.1959,PhysRevLett.95.137204,PhysRevLett.95.137205,PhysRevLett.97.026603,PhysRevLett.99.197202,PhysRevLett.107.236601,PhysRevLett.112.166601,PhysRevB.91.214405,Failde2021,failde2021orbital,PhysRevResearch.3.013058}. These advances are a consequence of the particular and favourable transport properties that these systems share and that have a direct implementation through graphene-like systems, topological insulators (TIs), Chern insulators (CIs) and Weyl semimetals \cite{Bhimanapati2015,Liu2015,Konig766,Zhang2009,Lieaaw5685,Deng895,Xue2018}. Graphene-like systems are intentionally mentioned because when using a common honeycomb lattice we usually need to introduce new terms in the Hamiltonian that the graphene structure does not take into account, that is, for instance, a spin-orbit coupling term and/or a staggered on-site potential that makes finite the Berry curvature \cite{PhysRevLett.95.146802,PhysRevLett.95.226801,PhysRevLett.97.036808}. At the same time, when working with TIs or CIs we need to take note of the dimensions of the space-time we are working and the intrinsic symmetries of the system \cite{PhysRevB.75.121306,PhysRevB.76.045302,PhysRevB.78.195125}. Generally, most of the obtained results focus on two-dimensional (2D) materials where the Abelian Berry curvature allows a simple determination of the system\textquotesingle s topology by means of the first Chern number $C$ \cite{PhysRevB.81.115407}. A generalization to higher dimensional systems is not 	immediate since this process involves dealing with non-Abelian terms in our calculations \cite{PhysRevB.78.195424,Baldomir2019}. 

Throughout different formalism, it has been shown how some properties of these systems such as the Berry curvature, density of states, orbital magnetic moment and energy behave when applying a magnetic field \cite{PhysRevLett.95.137204,PhysRevLett.99.197202,PhysRevLett.112.166601,PhysRevB.91.214405,failde2021orbital,PhysRevResearch.3.013058}. The last effort provides analytical expressions to show how these magnitudes depend on the external magnetic field \cite{failde2021orbital}. Thus, it can be seen how an external magnetic field modulates the shape of the Berry curvature maintaining its topological invariant. However, these calculations apply to any 2D non-zero Berry curvature material with a linear or quasi-linear dispersion law while 2D and 3D TIs thin-films usually present non-negligible parabolic terms in their spectra \cite{Konig766,Zhang2009}. To introduce such a term and study the orbital magnetization and susceptibility as well as other transport properties in an accurate way one needs to reach second-order corrections. This is done because some properties as the susceptibility or conductivities involve second derivatives in the energy with respect to the electromagnetic fields and the corrections become non-negligible at relatively low external fields due to the low energy nature of the relativistic models working on these materials \cite{PhysRevLett.112.166601,failde2021orbital,PhysRevResearch.3.013058}. On the other hand, while an exact solution in terms of Landau levels can be achieved for a constant mass in an external magnetic field this method is likely to result tortuous when considering other interactions compared with the other specified formalisms \cite{PhysRevB.81.195431}.

In this study, we introduce the role of the momentum dependent mass term in the corrections to the Dirac Hamiltonian  in presence of a magnetic field. Reaching second-order corrections to all the physical magnitudes involved, our calculations break the current scenario in which both topological trivial and non-trivial regimes has an equivalent orbital magnetic response. In this line, we observe that the trivial insulator has a limited response to this parameter while a non-trivial insulator is predicted to show an anomalous response which can be written in terms of the square of the quantum magnetic flux. As an immediate consequence, it is shown how the intrinsic orbital magnetism of topologically non-trivial Berry curvature systems can be enhanced by increasing the parabolic dependence of the energy spectrum. The present results are suitable to increase the knowledge of the topological phase transitions and  have a direct verification in the experimental field.

\begin{figure*}[t]
\includegraphics[scale=0.915]{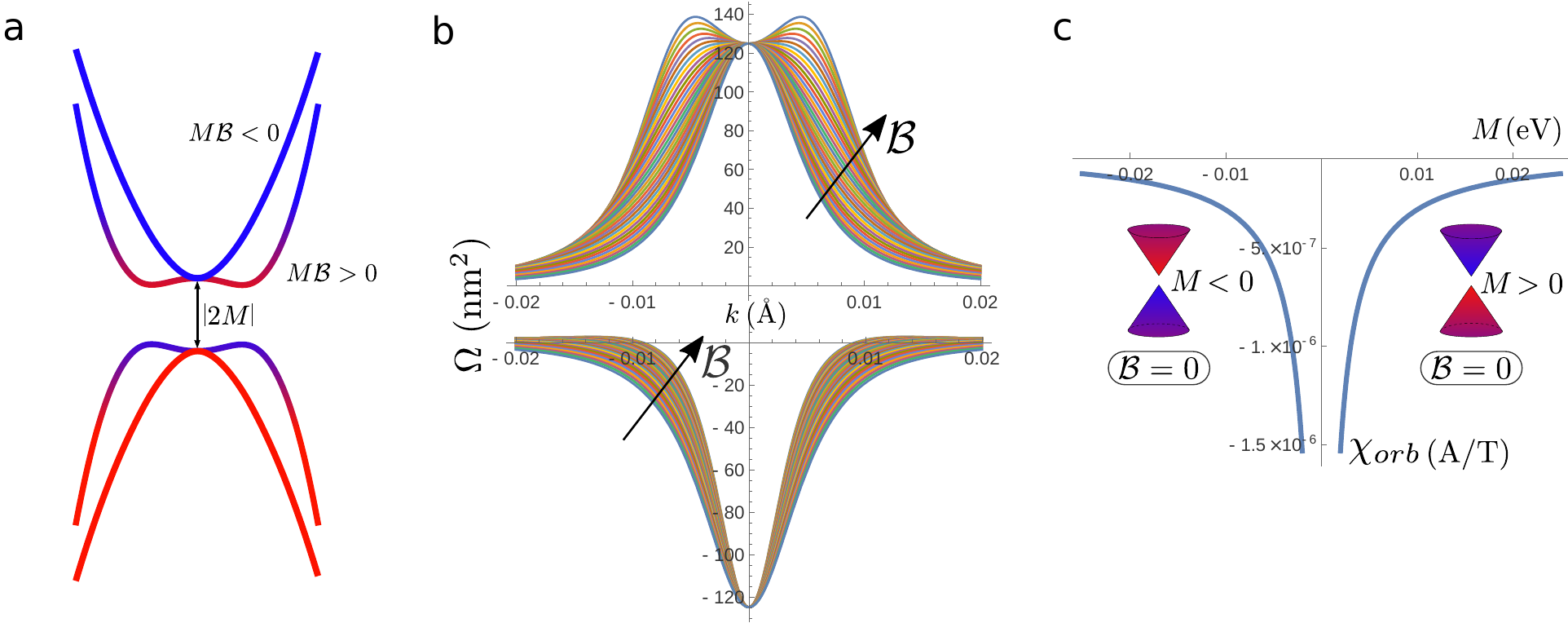}
\caption{\textbf{Two-dimensional features of low energy relativistic models.} \textbf{a}, Energy spectra of a two-dimensional Dirac Hamiltonian $H_{2D}$ for the trivial ($M>0$, $\mathcal{B}<0$) and non-trivial ($M<0$, $\mathcal{B}<0$) cases. The color gradient indicates the $z$ component of the k-dependent spin texture of each regime. In this configuration, the red color stands for spin down and blue for spin up \textbf{b}, Berry curvature of the conduction band of $H_{2D}$ for different values of $\mathcal{B}$ ranging from 0 to $-300$ eV\AA$^2$ in steps of 10 eV\AA$^2$. The arrows indicate the change in $\Omega$ when increasing $\mathcal{B}$. \textbf{c}, Zero-field orbital magnetic susceptibility at $\mathcal{B}=0$ (gaped Dirac dispersion) as a function of $M$.}
\end{figure*}
\section*{Results}
Our starting point is a solid-state version of the Dirac Hamiltonian used to describe 2D and 3DTIs \cite{Bernevig1757},
\begin{equation}
    H= M(\boldsymbol{k}) \beta + v_F(\boldsymbol{\alpha} \cdot \boldsymbol{p}) 
\label{Hamiltonian}
\end{equation}
where the Fermi velocity $v_F$ substitutes the speed of light $c$, $\beta$ and $\alpha_i\;(\sigma_i)$ are the Dirac (Pauli) matrices, $\boldsymbol{p}=\hbar \boldsymbol{k}$ the momentum of the particles and $M(k)=M-\mathcal{B}k^2$ a k-dependent diagonal term where $M$ and $\mathcal{B}$ depend directly on the on-site energies and second-neighbour hopping elements. This effective Hamiltonian allows a simple determination of the system\textquotesingle s topology, especially on its two-dimensional version where we can easily differentiate between the topological non-trivial and trivial regimes by looking to the relative sign between the parameters $M$ and $\mathcal{B}$. That is, it reduces to a Bernevig-Hughes-Zhang (BHZ) model with two non-interacting time-reversal copies ($M(k)\rightarrow-M(k)$) of a 2D Dirac Hamiltonian $H_{2D}=M(k)\sigma_z+v_F\, \boldsymbol{\sigma}\cdot \boldsymbol{p}$ \cite{Bernevig1757,PhysRevB.81.115407}. Thus, we have $M\mathcal{B}>0$ for the non-trivial regime and $M\mathcal{B}<0$ for the trivial one that fixes the integer Chern number $C$ to be $1$ or $0$ respectively. Notice that $\mathcal{B}$ is usually negative as the energy tends to grow when increasing $k$ and the energy spectrum of Eq\;\eqref{Hamiltonian} is given by $\xi=\pm\sqrt{(M-\mathcal{B}k^2)^2+\hbar^2v_F^2k^2}$. So essentially the topology depends on the sign of $M$, which for the particular case $M<0$ (inverted band structure) results non-trivial. In Fig.\;1a we can find a representation of the energy spectra and spin-configuration of the bands of both non-trivial and trivial regimes in $H_{2D}$. For the full Hamiltonian Eq.\;\eqref{Hamiltonian}, Dirac hyperbolae shall be plotted twice to take into account spin degeneracy.  A similar Hamiltonian can be obtained from a Kane-Mele model on a honeycomb lattice although with a different topological constraint for the $M\mathcal{B}$ parameters \cite{PhysRevLett.95.226801}.

\begin{figure*}[t]
\includegraphics[scale=1.0]{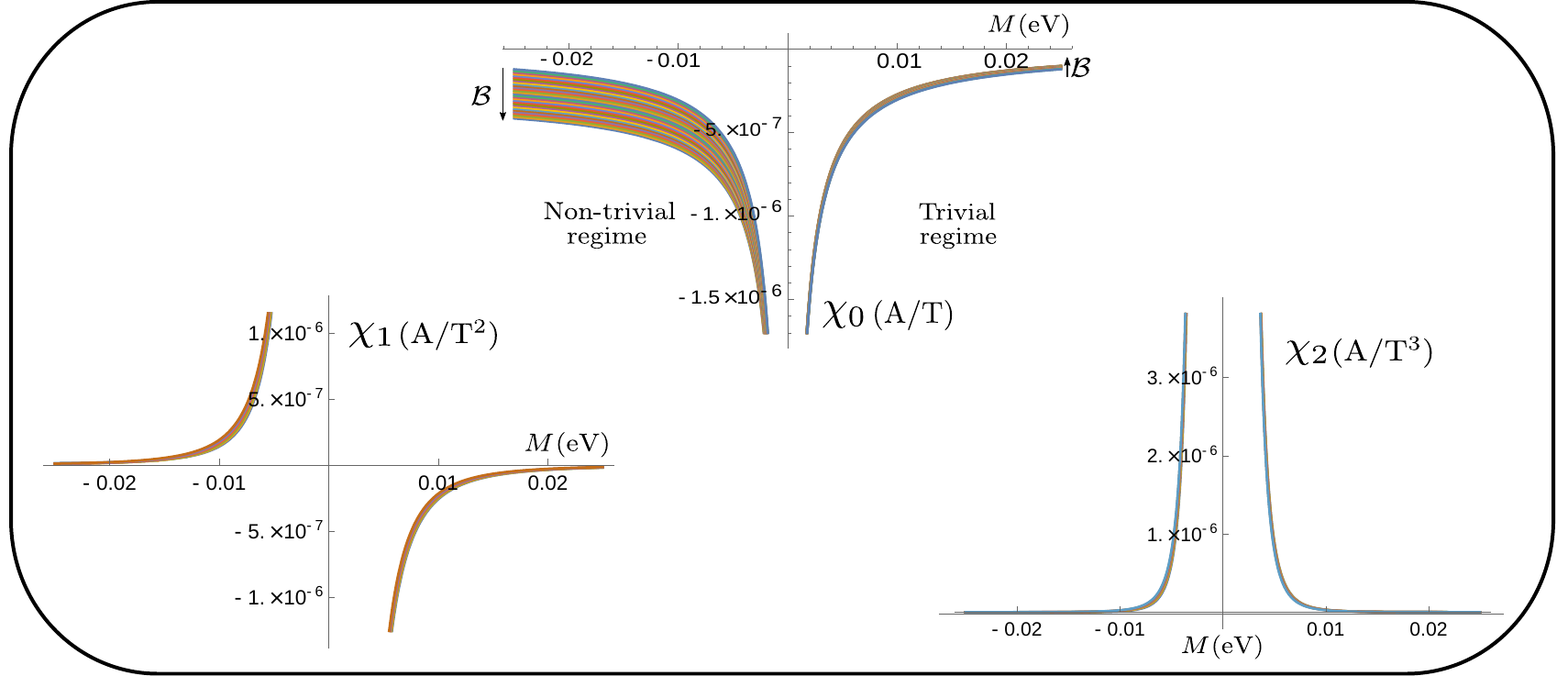}
\caption{\textbf{Orbital magnetic susceptibility in non-zero Berry curvature systems.} \textbf{a}, Zero-field orbital magnetic susceptibility of the trivial and non-trivial topological regimes. $\mathcal{B}$ ranges from $0$ to $-300$ eV\AA$^2$ and $v_F$ is set to $6\cdot 10^5$ m/s. In contrast to the current scenario displayed in Fig. 1c, the introduction of the parabolic dependence at large $k$ given by $\mathcal{B}$ determines a drastic change in the behaviour of both regimes  \textbf{b,c}, Linear ($\chi_1$) and quadratic ($\chi_2$) coefficients in the magnetic field of the orbital susceptibility.}
\end{figure*}

Introducing a perpendicular magnetic field $\boldsymbol{B}=B \boldsymbol{\hat{z}}$ on $H_{2D}$ causes a correction in the eigenstates of the systems that can be written as \cite{PhysRevLett.112.166601,failde2021orbital}

\begin{equation}
    \ket{n}\rightarrow \ket{n}+ \frac{eB}{(\xi^n-\xi^m)} \frac{1}{\hbar}\left( \partial_{k_y}\xi^m A_x^{mn}-  \partial_{k_x}\xi^m A_y^{mn}\right)\ket{m}
\label{corrections}
\end{equation}
where $-e$ is the electron charge, $A_j^{mn}=i\bra{m}\ket{\partial_{k_j}n}$, $\ket{n}$ and $\ket{m}$ are the eigenstates of the bands $n$ and $m$ with energy $\xi^n$ and $\xi^m$. These corrections represent the coupling effects between the magnetic field and Berry curvature of the bands which induce second-order corrections to the Berry potential, Berry curvature, particle\textquotesingle s velocity, density of states, orbital magnetic moment and energy \cite{failde2021orbital}. Thus,  for the Berry potential $\boldsymbol{A}^n=i\bra{n}\ket{\partial_{\boldsymbol{k}} n}$ we find from Eq.\;\eqref{corrections} that
\begin{equation}
    A_j^n \rightarrow A_j^n+2\Re A_j^{nm} \frac{\hbar \omega_z^n}{(\xi^n-\xi^m)}
    \label{potential}
\end{equation}
being $\omega_z^n=\epsilon_{zrs}\frac{eB_z}{\hbar}\frac{1}{\hbar}\partial_{k_r}\xi^m A^{mn}_s$ as it can be deduced by different formalism \cite{PhysRevLett.112.166601,failde2021orbital}. These expressions involve quite tedious calculations especially if we want to take into account the role of the parameter $\mathcal{B}$ in our Hamiltonian or deal with a particle-hole antisymmetric system. We are going to focus on the first case given that as the current literature indicate, there are no differences in the orbital magnetic susceptibility at zero-field $\chi_{orb}(B=0)$ between the trivial and non-trivial regimes when we consider a gaped Dirac dispersion \cite{PhysRevB.91.214405,PhysRevResearch.3.013058,failde2021orbital,Fukuyama,PhysRevB.81.195431}. This is quite impressive given that both regimes present clear differences in their curvatures and $\chi_{orb}$ is directly related with it (Fig 1b). Remember that at zero temperature $\chi_{orb}=-\partial^2 E/\partial B^2$ being $E=\int \frac{d\boldsymbol{k}}{(2\pi)^2}D \tilde{\xi}$ the density of energy, $D=1+\frac{e\boldsymbol{B}\cdot\boldsymbol{\tilde{\Omega}}}{\hbar}$ the modified density of states, $\tilde{\Omega}$ the modified Berry curvature and $\tilde{\xi}=\xi\left(1-\frac{e\boldsymbol{B}\cdot\boldsymbol{\Omega}}{\hbar}+\mathcal{O}(B^2)\right)$ the corrected energy resulting from the application of the magnetic field. Then, one would expect to observe some differences although under linear dispersion conditions there are not (Fig 1b). Introducing $\mathcal{B}$ and after some algebra, it can be shown that the corrections of the Berry curvature $\Omega^n=-2Im\bra{\partial_{k_x}n}\ket{\partial_{k_y}n}=\boldsymbol{\nabla}\times \boldsymbol{A}$, obtained by applying Eq.\;\eqref{corrections} or Eq.\;\eqref{potential} to the eigenstates of $H_{2D}$, are given by
\begin{equation}
    \tilde{\Omega}=\Omega+2\frac{eB\Omega}{\hbar}\Omega\left(\frac{M-\mathcal{B}k^2}{(M+\mathcal{B}k^2)^2}F(k^2)-
    \frac{\hbar^2v_F^2k^2}{(M+\mathcal{B}k^2)^2}G(k^2)\right)
    \label{curvature}
\end{equation}
being $\xi=\pm\sqrt{(M-\mathcal{B}k^2)^2+\hbar^2v_F^2k^2}$, $\Omega=- \hbar^2v^2(M+\mathcal{B}k^2)/(2\xi^3)$ and
\begin{equation*}
    F(k^2)=M-\frac{\xi^2\mathcal{B}}{\hbar^2v_F^2}-\frac{\mathcal{B}(M-\mathcal{B}k^2)(M+\mathcal{B}k^2)}{\hbar^2v_F^2}
\end{equation*}
\begin{equation*}
    G(k^2)=1-6\mathcal{B}\frac{M-\mathcal{B}k^2}{\hbar^2v_F^2}+6\mathcal{B}^2\frac{(M-\mathcal{B}k^2)^2}{\hbar^4v_F^4}-2\mathcal{B}^2\frac{\xi^2}{\hbar^4v_F^4}
\end{equation*}
Eq.\;\eqref{curvature} generalizes the results obtained for linear dispersions introducing $\mathcal{B}$ which is essential to define properly the Chern number $C=1/(2\pi)\int\Omega d\boldsymbol{k}$ as an integer and distinguish the topological regime from the trivial one \cite{failde2021orbital,PhysRevB.81.115407}. Numerically, it can be tested that the Chern number of the system is fixed for any given value as long as we maintain the sign of $M\mathcal{B}$ invariant. 

\begin{figure*}[t]
\includegraphics[scale=0.96]{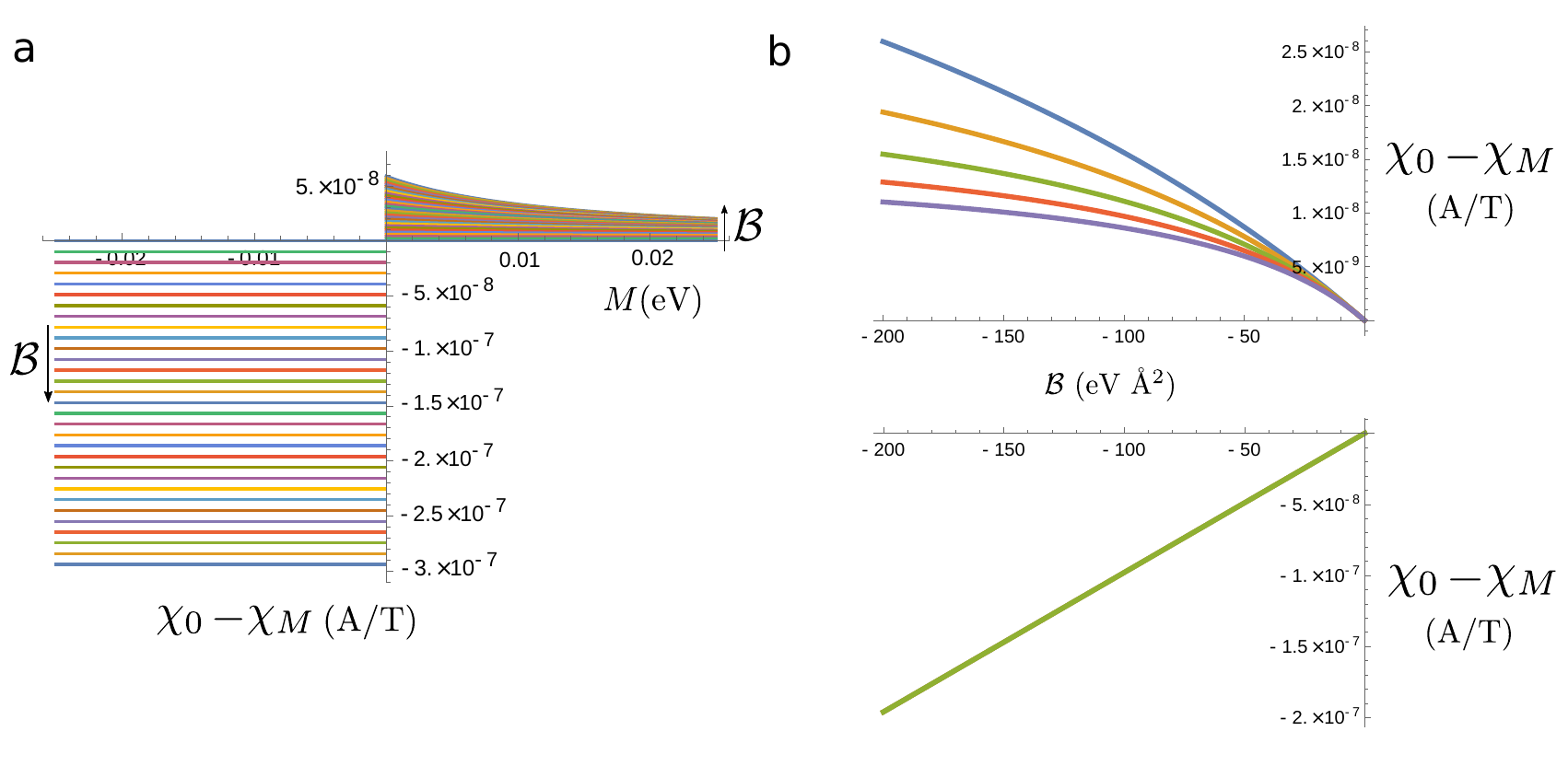}
\caption{\textbf{Normalized zero-field orbital magnetic susceptibility.} \textbf{a}, $\chi_0-\chi_M$ vs $M$ for different $\mathcal{B} \in [-300,0]$\AA.   \textbf{b}, $\chi_0-\chi_M$ vs $\mathcal{B}$ for different values of $M$ ranging from $10$ (blue line) to $50$\,meV (purple line). The upper panel represents the behaviour of the normalized zero-orbital  susceptibility in a trivial insulator $M\mathcal{B}<0$ whereas the lower panel stands for a topologically non-trivial insulator ($M\mathcal{B}>0$) . This latter case exhibits no dependence on the gap giving a linear diamagnetic contribution to the zero-orbital susceptibility.  }
\end{figure*}
Having $\tilde{\Omega}$ and hence $D$, we only need to calculate the second-order corrections for the energy induced by the perpendicular magnetic field $B$
\begin{equation}
    \tilde{\xi}=\xi-mB+\frac{1}{2} \xi \left(\frac{eB\Omega}{\hbar} \right)^2 \frac{\hbar^2 v_F^2k^2}{M+\mathcal{B}k^2}\left(1-2\mathcal{B}\frac{(M-\mathcal{B}k^2)}{\hbar^2v^2}\right)^2
\end{equation}
and then it is straightforward to show how the orbital magnetization at zero-temperature $\mathcal{M}_{orb}=-\partial E/\partial B$ is a function of the magnetic field $B$ given by $\mathcal{M}_{
orb}=\mathcal{M}_1 B+\mathcal{M}_2 B^2+\mathcal{M}_3 B^3$. Or equivalently, we have an orbital magnetic susceptibility $\chi_{orb}=\chi_0+\chi_1 B+\chi_2 B^2$ where
\begin{equation}
    \chi_0=\frac{e^2v_F^2(-1+(1-2rs)\beta+10(1+rs)\beta^2)}{6\pi(1-4\beta)\abs{M}}
\end{equation}
\begin{equation}
    \chi_1=-\frac{3e^3\hbar v_F^4}{64 \pi M^3 (1-4\beta)^{3/2}} \Lambda(\beta)
\end{equation}
\begin{equation}
    \chi_2=\frac{e^4\hbar^2v_F^6}{420\pi \abs{M}^5 (1-4\beta)^3}\Theta (\beta)
\end{equation}
We defined $\beta=M\mathcal{B}/\hbar^2v_F^2$, $s=sgn(M)$ and $r=sgn(\mathcal{B})$. A complete expansion of the functions $\Lambda(\beta)$ and $\Theta(\beta)$ can be found in the Methods section. Notice that $\chi_{orb}$ is given in units of amperes (A) per tesla (T) being typical to represent it as a dimensionless quantity by using a normalization parameter $\chi_n$ as the inverse of effective magnetic permeability $\mu$ in two-dimensions. That is, $e^2av_F/(6\pi^2\hbar)$ for the particular case of a honeycomb lattice with lattice constant $a$ \cite{PhysRevB.91.214405,PhysRevResearch.3.013058}. In Fig. 2 we can see that $\chi_2$ is practically paramagnetic for any value of $M$ and $\mathcal{B}$, although a detailed inspection reveals a paramagnetic region defined by the condition 
$4M\mathcal{B}>\hbar^2v_F^2$ ($\beta>1/4$) for the non-trivial topological regime (Methods). This term is liable to play an important role only for tiny gaps and/or high magnetic fields where corrections due to Zeeman effects should also be considered. That is, for a typical gap $\abs{2M}=50$meV, $v_F=6\cdot 10^5$ m/s and a field $B=1$T whose equivalent energy ($\hbar\omega$) does not exceed this gap we have that $\abs{\chi_2/\chi_0}\approx10^{-3}$ and $\abs{\chi_2/\chi_1}\approx10^{-2}$. Attending to $\chi_1$, its contribution to the orbital magnetization and susceptibility is opposite for both regimes resulting in paramagnetic for the topological non-trivial and diamagnetic for the normal state when $B$ is positive. Compared to $\chi_0$ the ratio between both is close to $10^{-1}$ under the same prior conditions and hence become to be non-negligible at relative low-magnetic fields depending on the band gap $2M$ of the system. Its dependence on $\mathcal{B}$ can be explored in more detail in the Methods. Thus, besides it is evident that $\chi_2$ and $\chi_1$ have a weaker dependence on this parameter compared with $M$ (Fig. 2b and c), for a fixed gap both regimes shows a very different response, with a nearly constant dependency for the trivial regime and strong for the topological one. Finally, note that for time-reversal symmetric systems such as Eq.\;\eqref{Hamiltonian} the contribution of the two Chern species cancels each other and $\chi_1$ will be zero while the even pairs in the magnetic field for the susceptibility appear with a spin-degeneracy factor $g=2$.

A significant result is obtained when analyzing the zero-field orbital susceptibility $\chi_0$. A first-look to Fig. 2a indicates that there are clear differences with the current scenario in which $\chi_0$ manifest the same behaviour for the non-trivial and trivial cases (Fig 1b). However, we can go further by subtracting the common \textit{background} term at zero $\mathcal{B}$ displayed in Fig 1c from Eq.\;(7). This factor is obtained by different methods and it is equal to $\chi_M=-\frac{e^2v^2}{6\pi\abs{M}}$ \cite{PhysRevB.91.214405,PhysRevResearch.3.013058,failde2021orbital,Fukuyama,PhysRevB.81.195431}.  In this way, we find that it appears a continuum spectrum for $\chi_0-\chi_M$ with a ratio $\frac{5}{6\pi} \frac{e^2}{\hbar^2}$ given in fundamental units when plotted versus $\mathcal{B}$ (Fig. 3). This unexpected behaviour belongs uniquely for the topological regime while for the trivial we have a non-homogeneous dependence on $\mathcal{B}$. In general formulae, we can write the zero-orbital magnetic susceptibility for both topological non-trivial 
\begin{equation}
    \chi_0=-\frac{e^2v^2}{6\pi \abs{M}}+\frac{5e^2}{6\pi\hbar^2}\mathcal{B} \qquad (M<0,\mathcal{B}<0)
\label{non-trivial}
\end{equation}
and trivial phases
\begin{equation}
    \chi_0=-\frac{e^2v^2 (\hbar^2v_F^2-3M\mathcal{B})}{6\pi \abs{M}(\hbar^2v_F^2-4M\mathcal{B})} \qquad (M>0,\mathcal{B}<0)
\label{trivial}
\end{equation}
by introducing in the same expression the principal parameters $M$, $\mathcal{B}$ and $v_F$ for the topology and transport which are directly related with the different hopping elements. Thus, roughly speaking for the simplest case of a square lattice with complex spin-dependent first neighbour hopping  $it_1$ and staggered potential for the on-site energies $\xi_v$ and second neighbour hopping $t_2$ we can reach for the topological regime that
\begin{equation}
    \mu\chi_0\propto-C_1 \frac{t_1}{\abs{\xi_v+4t_2}}-C_2\frac{t_2}{t_1}
\end{equation}
obtaining from the tight-binding that $v_F=2 at_1/\hbar$, $M=\xi_v+4t_2$, $\mathcal{B}=-2t_2a^2$, $\mu^{-1}\propto e^2v_Fa/\hbar$. Numerical factors accompanying these relations will depend on the crystal structure of the system which determine the coefficients $C_1$ and $C_2$.  

We can also observe by computing the large $\mathcal{B}$ limit ($M\mathcal{B}>>\hbar^2v_F^2$)  that $\chi_0$ grows linearly for the topological regime while it saturates in a finite value $-\frac{e^2v^2}{8\pi M}$ ($M>0$) for the trivial insulator with a softer dependence on this parameter (Fig.\;3b). The role of $\mathcal{B}$ opens a new possibility for tuning the orbital magnetization and susceptibility in topological materials breaking the current scenario in which we need to employ materials with a tiny gap for maximizing the response. This could be experimentally unpleasant given that then you are limited to work with a small frame of temperature and/or magnetic fields. Notice that the orbital susceptibility, also known as geometrical, is dominant only inside the band gap and decays to zero  as the Fermi level enters the conduction and valence bands \cite{PhysRevB.91.214405}. In other words, a more robust parabolic dependence in the energy spectra causes an enhanced orbital magnetic response in topological materials. 

The present results have another direct experimental consequence. On the one hand, directly from Eq\;\eqref{trivial} it is straightforward to note that the dependence on $\mathcal{B}$ in the trivial regime seems to be weak, as one can verify it by computing the ratio $(\hbar^2v_F^2-3M\mathcal{B})/(\hbar^2v_F^2-4M\mathcal{B})$ under the high Fermi velocity and small gap conditions that these systems feature. Typically, these are in the range of $10^5-10^6$ m/s for the velocity while the gap is in the order of meV \cite{Bernevig1757,Elias2011,Zhang2009}.  In contrast, a topological non-trivial insulator shows a ratio $\propto \frac{e^2}{\pi \hbar^2}$ independent of such parameters which can be related directly with the quantum magnetic flux $\Phi_0=h/(2e)$. Thus, varying $\mathcal{B}$ we expect a change in the orbital magnetic susceptibility $\Delta \chi_0 \propto \pi/\Phi_0^2$. This result seems to connect perfectly with the concept of orbit for the topological electrons developed in 2D and 3DTI thin films by translating the topological information  contained in the Berry curvature into a topological effective field $b$ \cite{Failde2021}. This theory relies on the concept of orbit for the topological electrons, for which, the topological quantization by means of the Chern number allows to associate a quantized flux to each orbit and hence an effective magnetic field $b$ and area $\Delta S$ to these electrons.
\begin{equation}
    b\approx\frac{(2)\mathbbm{m}^2 v_F^2}{\hbar e}   \qquad  \Delta S\approx\frac{h \hbar}{(2) \mathbbm{m}^2 v_F^2}
\label{emergent}
\end{equation}
Here $\mathbbm{m}$ is the effective mass and the factor (2) vanishes when redefining the orbital magnetic moment \cite{failde2021orbital}. That is, matching $b$ with the solid-state version of the Schwinger critical field \cite{PhysRevD.72.105004}. In summary, it can be defined a magnetic field $b$ from the Berry curvature on each band of a TI or Chern insulator with net value zero in presence of time-reversal symmetry.  Mathematically, it is straightforward to connect these results with the orbital magnetic susceptibility given that by its proper definition $b$ and $\Delta S$ define a quantized flux $\Phi=h/e $ associated to each band. So, this magnetic flux should be reflected as is in the system's response to an external magnetic field. This must occur when one band is fully filled, i.e., when the Fermi level lies in the band gap and the orbital susceptibility is dominant. Physically, by looking at the energy dispersion it is obvious that varying $\mathcal{B}$ makes it more localized/delocalized in the k-space and modifies the effective mass of the particles. Given that the topology is the same ($sgn(M\mathcal{B})$) remains invariant), this would change the effective area of the electrons $\Delta S$ and also the field $b$ associated to each band to keep its flux invariant inducing a singular magnetic response that needs to be related to this flux.

Further corrections can be taken into account implying that the numerical factor accompanying $\pi/\Phi_0^2$ in the second term of Eq.\;\eqref{non-trivial} should experiment changes. Notice that by considering $\mathcal{B}$ in our Hamiltonian we are introducing a k-dependent mass term in our system. So the perpendicular magnetic field should also give a diagonal contribution in the perturbed Hamiltonian that could be introduced in Eq.\;\eqref{corrections}. This would lead to more accurate expressions but from the physical point of view, this would also imply the necessity to consider Zeeman corrections. Nevertheless, if one performs the simple calculation associated with first-order corrections of this term to the energy it is easy to show that $\bra{n}\Delta H_z\ket{n}=\frac{eB}{2\hbar}\mathcal{B}$ being $\Delta H_z=\frac{eB}{\hbar}\mathcal{B}\sigma_z(k_x y-k_y x)$. Comparing this factor with the first-order term $-\boldsymbol{m\cdot B}$ we can verify that at low energy $\frac{eB\mathcal{B}/\hbar}{eB\xi\Omega/\hbar}\approx\frac{M\mathcal{B}}{\hbar^2v^2}$ is again negligible for a wide range of parameters that typically characterize these materials.
\section*{Conclusion and outlook}
All materials respond as paramagnetic or diamagnetic under an applied external magnetic field although their physical origins are very different. For example, diamagnetism depends mainly on the atomic number and the orbital electron radius, while paramagnetism strongly depends on the temperature and magnetic momenta of the ground state within the first order approach. This last condition can be overcome by introducing higher order perturbative terms as occurs with van Vleck paramagnetism when also considering the magnetization of excited states. However, in 2D materials with non-zero Berry curvature, new orbital magnetizations emerge from this property. In this context, as we show in Figure 2 and 3, we found new terms in the orbital magnetic susceptibility of the topological non-trivial regime which are susceptible to be measured and open a new form to modulate the intrinsic orbital magnetism of topological materials. 

Concretely, we theoretically analyzed the orbital dynamics of 2D non-zero Berry curvature systems by introducing the natural $k$ dependence in the energy dispersion and mass that the Dirac electrons have in condensed matter. Given the crucial role of this term for a proper definition of the topological invariant Chern number, the analytical expressions which results from this perturbative analysis of the Dirac Hamitonian in presence of a perpendicular magnetic field are able to disentangle the topological orbital effects besides extending the previous results using linear dispersion to a broader variety of compounds. The new scenario which arises involve drastic differences in the orbital magnetic response of the topological trivial and non-trivial regimes. Thus, while a trivial insulator is predicted to have a limited dependence on $\mathcal{B}$ based on the common values that these materials have, its topological counterpart features a unexplored term in the zero-field orbital magnetic susceptibility which is given in fundamental units and is inversely proportional to the square of the quantum magnetic flux. The results obtained go further in the study singular physics of these materials approaching it to the quantum phase transition phenomenology and enables different paths to modulate the orbital magnetization and susceptibility in these materials. As well, they establish the basis to treat directly other non-linear thermoelectric and magnetoelectric properties of non-zero Berry curvature systems. 

\section{Methods}
\textbf{Magnetic field dependent terms in the orbital susceptibility} Starting from the eigenstates of $H_{2D}=M(k)\sigma_z+v_F\, \boldsymbol{\sigma}\cdot \boldsymbol{p}$ it is possible to reproduce the results obtained after some algebra. 
\begin{equation*}
    \ket{\pm}=\frac{1}{\sqrt{2}}\left[\begin{array}{c} \sqrt{1\pm\frac{M({ k})}{\xi}} \\ \pm e^{i\phi} \sqrt{1\mp \frac{M({ k})}{\xi}} \end{array}\right]
\end{equation*}
We labeled them as $\ket{\pm}$, being $\ket{+}$ the eigenstate with energy $\xi$ (conduction band) and $\ket{-}$ that with negative energy (valence band). To perform the calculations, we need to consider the following partial derivatives of the eigenstates
\begin{equation*}
    \ket{\partial_{kx} +}=\frac{1}{\sqrt{2}}\left[\begin{array}{c} \frac{\Gamma_{kx}}{2\sqrt{1+\frac{M({ k})}{\xi}}} \\ \frac{-ik_y}{k^2}e^{i\phi} \sqrt{1-\frac{M({ k})}{\xi}}-\frac{\Gamma_{kx}e^{i\phi}}{2\sqrt{1-\frac{M({ k})}{\xi}}} \end{array}\right]
\end{equation*}\hspace{5pt}
\begin{equation*}
    \ket{\partial_{ky} +}=\frac{1}{\sqrt{2}}\left[\begin{array}{c} \frac{\Gamma_{ky}}{2\sqrt{1+\frac{M({ k})}{\xi}}} \\ \frac{ik_x}{k^2}e^{i\phi} \sqrt{1-\frac{M({ k})}{\xi}}-\frac{\Gamma_{ky}e^{i\phi}}{2\sqrt{1-\frac{M({ k})}{\xi}}} \end{array}\right]
\end{equation*}
as well as the ones corresponding to the negative solution which are involved in some matrix elements which are worthy to be mentioned. Here $\Gamma_{k_i}=-\frac{\hbar^2v_F^2 (M+Bk^2)}{\xi^3}k_i$. The calculations are done employing an axial gauge for the vector potential $\boldsymbol{A}=(-By/2,Bx/2,0)$ which simplifies them by removing the phase factor $e^{i\phi}$ present in our eigenstates, where $\phi=arctan(k_y/k_x)$. Some useful relations are
\begin{equation*}
  i\left(k_y \bra{-}\ket{\partial_{k_x}+}- k_x \bra{-}\ket{\partial_{k_y}+}\right)=-\frac{\hbar v_F k}{\xi}
\end{equation*}\hspace{5pt}
\begin{equation*}
  i\left(\bra{\partial_{k_x}-}\ket{\partial_{k_y}+}- \bra{\partial_{k_y}-}\ket{\partial_{k_x}+}\right)=-\Omega^+\frac{M(k)}{\hbar v_F k}
\end{equation*}
where $\Omega^+$ stands for the Berry curvature of the conduction band. With these elements we can finally give an expression for the magnitudes previously specified. Thus, in general notation the coefficients of the orbital susceptibility result
\begin{widetext}
\begin{equation*}
    \chi_0=\frac{e^2v_F^2(-1+(1-2rs)\beta+10(1+rs)\beta^2)}{6\pi(1-4\beta)\abs{M}}
\end{equation*}
\begin{equation*}
    \chi_1=-\frac{3e^3\hbar v_F^4 \left( \sqrt{1-4\beta} \; \left( 1-6\beta-6\beta^2(-1+2\beta)\right)+4\beta^3(-2+3\beta)\left(2 \arctanh{\left(\frac{-1+2\beta}{\sqrt{1-4\beta}}\right)}-\ln\left(-\frac{\beta^2}{\sqrt{1-4\beta}}\right)+\ln\left(\frac{\beta^2}{\sqrt{1-4\beta}} \right)\right)   \right)}{64 \pi M^3 (1-4\beta)^{3/2}} 
\end{equation*}
\begin{equation*}
    \chi_2=\frac{e^4\hbar^2v_F^6 \left( 1-14\beta+68\beta^2-120\beta^3+128(1+rs)\beta^5-512(1+rs)\beta^6+1024(1+rs)\beta^7\right)}{420\pi \abs{M}^5 (1-4\beta)^3}
\end{equation*}
\end{widetext}
It is important to note that $\chi_1$ is real for any value of $\beta$ and the imaginary factors arising when $\beta>1/4$ in both terms of the fraction cancel. A numerical representation of $\chi_1$ vs $\beta$ can be found in Extended Data Fig. 1.  The particular case $\beta=1/4$ corresponds to the condition $4M\mathcal{B}=\hbar^2v^2$ which makes the energy dispersion to be fully parabolic $\xi=\pm (M+Bk^2)$. Despite all terms contain a factor $(1-4\beta)$ in their denominators it can be checked that no divergences occur since the numerators also have the same root for positive $\beta$, i.e., for $M\mathcal{B}>0$. By computing the previous formulae for $\chi_1$ and $\chi_2$ it can be shown their weak dependence on $\mathcal{B}$ which results negligible for a wide range taking into account the different magnitude between these terms and the zero-field term $\chi_0$ (Extended Data Fig.\;1). This fact is more marked for the trivial regime ($\beta<1$) in all terms of the orbital magnetic susceptibility and more corrections might be expected to reduce the $\mathcal{B}$ contribution increasing the discrepancies between both topological trivial and non-trivial regimes. Notice that in the Extended Data the coefficients are normalized by their value at zero $\mathcal{B}$, so they still present a divergent behaviour when closing the band gap.
\setcounter{figure}{0}  
\renewcommand{\figurename}{\textbf{Extended Data Fig.}}
\begin{figure}[t]
\includegraphics[scale=0.92]{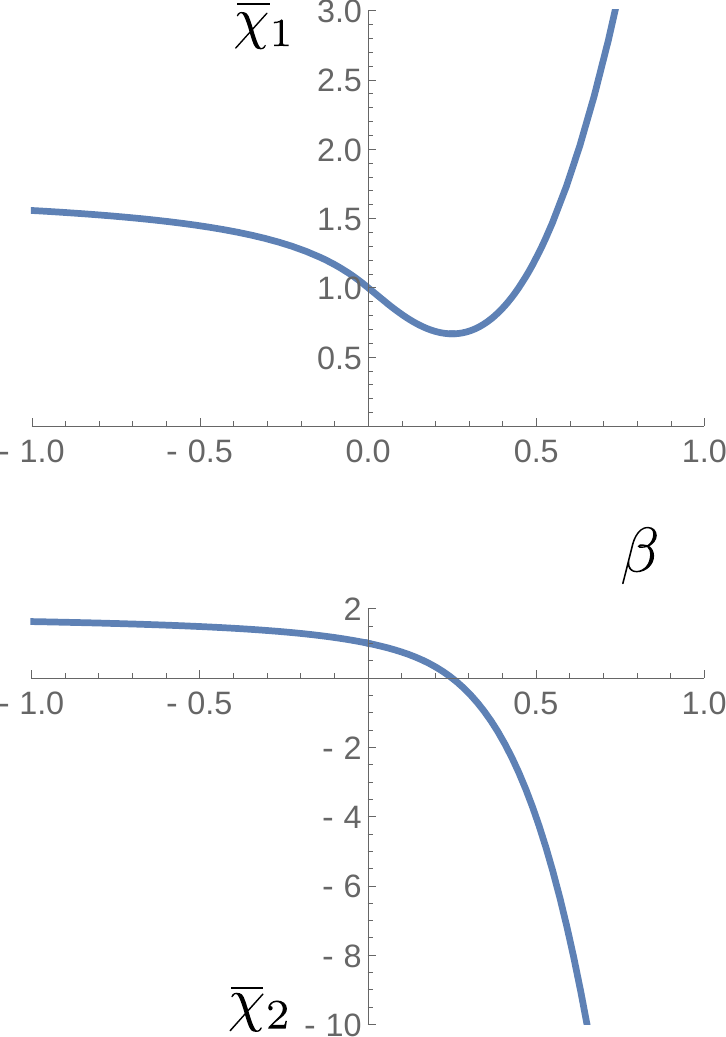}
\caption{\textbf{$\mathcal{B}$ dependence of the $\chi_1$ and $\chi_2$ coefficients.} Normalized coefficients $\bar{\chi}_1=\chi_1/\chi_1(\mathcal{B}=0)$ and $\bar{\chi}_2=\chi_2/\chi_2(\mathcal{B}=0)$ vs the dimensionless parameter $\beta=\frac{M\mathcal{B}}{\hbar^2v_F^2}$. Just as the zero-field term $\chi_0$, $\chi_1$ and $\chi_2$ also show clear differences between the trivial and non-trivial insulating  regimes. }
\end{figure}

\section{Data availability}
All data that support the plots within this paper and other findings of this study are available from the corresponding authors upon reasonable request.

\section{Acknowledgements}
Authors acknowledge to PID2019-104150RB-I00, AEMAT ED431E 2018/08 and the MAT2016-80762-R projects for financial support. We thank Juan Manuel Fa\'ilde for helpful discussions. Authors acknowledge to CESGA for computational facilities. 

\section{Author contributions}
D.F. conceived the problem. D.F. and D.B. made the calculations and wrote the manuscript.
\section{Competing interests}
The authors declare no competing interests.

\nocite{*}
\providecommand{\noopsort}[1]{}\providecommand{\singleletter}[1]{#1}%
\end{document}